\documentclass[prd,reprint,nofootinbib,notitlepage,aps,tightenlines,
preprintnumbers,amsmath,amssymb,showpacs,superscriptaddress]{revtex4-1}

\usepackage[dvipdfmx]{graphicx}
\usepackage[hyperindex,breaklinks]{hyperref}
\usepackage{graphicx,epstopdf,amsmath,amsfonts,amssymb,color,comment,cancel,subfigure}

\usepackage{tikz}

\usepackage{chngcntr}
\counterwithin{paragraph}{subsection}

\def\be{\begin{equation}}
\def\ee{\end{equation}}
\def\ba{\begin{eqnarray}}
\def\ea{\end{eqnarray}}
\def\ge{\mathrel{\raise.3ex\hbox{$>$\kern-.75em\lower1ex\hbox{$\sim$}}}}
\def\la{\mathrel{\raise.3ex\hbox{$<$\kern-.75em\lower1ex\hbox{$\sim$}}}}

\def\thesection{\arabic{section}}
\def\theequation{\arabic{equation}}
\def\simgt{\mathrel{\raise.3ex\hbox{$>$\kern-.75em\lower1ex\hbox{$\sim$}}}}
\def\simlt{\mathrel{\raise.3ex\hbox{$<$\kern-.75em\lower1ex\hbox{$\sim$}}}}

\newcommand{\nc}{\newcommand}

\nc{\gone}{\bar g_{\pi NN}^{(1)}}
\nc{\gzero}{\bar g_{\pi NN}^{(0)}}
\nc{\al}{\alpha}
\nc{\ga}{\gamma}
\nc{\de}{\delta}
\nc{\ep}{\epsilon}
\nc{\ze}{\zeta}
\nc{\et}{\eta}
\nc{\ka}{\kappa}
\nc{\rh}{\rho}
\nc{\si}{\sigma}
\nc{\ta}{\tau}
\nc{\up}{\upsilon}
\nc{\ph}{\phi}
\nc{\ch}{\chi}
\nc{\ps}{\psi}
\nc{\om}{\omega}
\nc{\Ga}{\Gamma}
\nc{\De}{\Delta}
\nc{\La}{\Lambda}
\nc{\Si}{\Sigma}
\nc{\Up}{\Upsilon}
\nc{\Ph}{\Phi}
\nc{\Ps}{\Psi}
\nc{\Om}{\Omega}
\nc{\ptl}{\partial}
\nc{\del}{\nabla}
\nc{\ov}{\overline}
\nc{\newcaption}[1]{\centerline{\parbox{15cm}{\caption{#1}}}}
\nc{\us}{U(1)$_S$}
\nc{\Rg}{$R_{\gamma\gamma}$}
\nc{\bbnu}{\beta\beta_{0\nu}}

\def\beq{\begin{equation}}
\def\eeq{\end{equation}}
\def\bmat{\begin{displaymath}}
\def\emat{\end{displaymath}}
\def\bear{\begin{eqnarray}}
\def\eear{\end{eqnarray}}
\def\ba{\begin{eqnarray}}
\def\ea{\end{eqnarray}}
\def\bery{\begin{array}}
\def\ery{\end{array}}
\def\bit{\begin{itemize}}
\def\eit{\end{itemize}}
\def\ben{\begin{enumerate}}
\def\een{\end{enumerate}}
\def\btab{\begin{tabular}}
\def\etab{\end{tabular}}
\def\btbl{\begin{table}}
\def\etbl{\end{table}}
\def\bfig{\begin{figure}[htb]}
\def\efig{\end{figure}}
\def\bpic{\begin{picture}}
\def\epic{\end{picture}}

\def\ga{\mathrel{\raise.3ex\hbox{$>$\kern-.75em\lower1ex\hbox{$\sim$}}}}
\def\la{\mathrel{\raise.3ex\hbox{$<$\kern-.75em\lower1ex\hbox{$\sim$}}}}
\def\gappeq{\mathrel{\rlap {\raise.5ex\hbox{$>$}}
{\lower.5ex\hbox{$\sim$}}}}
\def\lappeq{\mathrel{\rlap{\raise.5ex\hbox{$<$}}
{\lower.5ex\hbox{$\sim$}}}}

\def\gyr{{\rm \, G\kern-0.125em yr}}
\def\mev{{\rm \, Me\kern-0.125em V}}
\def\gev{{\rm \, Ge\kern-0.125em V}}
\def\tev{{\rm \, Te\kern-0.125em V}}

\def\slash#1{\rlap{\hbox{$\mskip 1 mu /$}}#1}%

\def\lsim{\mathrel{\rlap{\lower4pt\hbox{\hskip1pt$\sim$}}
    \raise1pt\hbox{$<$}}}                
\def\gsim{\mathrel{\rlap{\lower4pt\hbox{\hskip1pt$\sim$}}
    \raise1pt\hbox{$>$}}}                

\definecolor{jazzberryjam}{rgb}{0.65, 0.04, 0.37}        

\begin{document}
 
\title{Electric Dipole Moments From Dark Sectors}

\author{Shohei Okawa}
\affiliation{Department of Physics and Astronomy, University of Victoria, 
Victoria, BC V8P 5C2, Canada}

\author{Maxim Pospelov}
\affiliation{Department of Physics and Astronomy, University of Victoria, 
Victoria, BC V8P 5C2, Canada}
\affiliation{Perimeter Institute for Theoretical Physics, Waterloo, ON N2J 2W9, 
Canada}

\author{Adam Ritz}
\affiliation{Department of Physics and Astronomy, University of Victoria, 
Victoria, BC V8P 5C2, Canada}

\date{May 2019}

\begin{abstract}

We examine the sensitivity of electric dipole moments (EDMs) to new $CP$-violating physics in a hidden (or dark) sector, neutral under the Standard Model (SM) gauge groups, and coupled via renormalizable portals. In the absence of weak sector interactions, we show that the electron EDM can be induced purely through the gauge kinetic mixing portal, 
but requires five loops, and four powers of the kinetic mixing parameter $\epsilon$. 
Allowing weak interactions, and incorporating the Higgs and neutrino portals, we show that the leading contributions to $d_e$ arise at two-loop order, with the main source of $CP$-violating being in the interaction of dark Higgs and heavy singlet neutrinos. In such models, EDMs can provide new sensitivity to portal couplings that is complementary to direct probes at the intensity frontier or high energy colliders.

\end{abstract}
\maketitle

\section{Introduction}
\label{sec:intro}

Experiments at the high energy frontier of particle physics, namely at the LHC, have confirmed the Standard Model (SM) to high precision with no compelling evidence thus far for new physics. In turn, some of the strongest empirical evidence for physics beyond the SM, neutrino mass and dark matter, does not necessarily point to an origin at short distances. As a consequence, significant attention has been paid to models of physics involving hidden (or dark) sectors, as explanations of these empirical deficiencies of the SM. 

Dark sector models may involve new degrees of freedom with a mass well below the electroweak scale. The primary assumption is that all new fields are neutral under SM symmetries, implying in particular that the chiral electroweak $SU(2)_L\times U(1)_Y$ structure of the SM is maintained. The effective Lagrangian at the electroweak scale then takes the form
\be
 {\cal L}_{\rm NP} ={\cal L}_{\rm IR} + \sum_{d\geq 5} \frac{1}{\La_{\rm UV}^{d-4}} {\cal O}_{d},
\ee
with new UV physics described by a series of higher dimensional operators ${\cal O}_d$ constructed from SM fields. Notably, new IR physics contained in ${\cal L}_{\rm IR}$ is highly constrained by this effective field theory (EFT) framework, as it must be UV-complete. The simplest SM-neutral IR hidden sector allows only new scalars $S_i$, neutral fermions $N_i$ and/or new $U(1)'$ gauge boson(s) $A'_\mu$ \cite{Holdom:1985ag,*Foot:1991kb,*Foot:1991bp,*Pospelov:2007mp,*Batell:2009di} to mediate direct interactions with the SM. Indeed, the only renormalizable, and thus UV-complete, interactions (or portals) for these fields with the SM can be written in the form
\be
\label{portal}
 {\cal L}_{\rm IR} = \ep B^{\mu\nu}F'_{\mu\nu} - (AS + \lambda S^2)H^\dagger H - Y_N LHN + {\cal L}_{\rm hid},
\ee
possibly generalized to include multiple copies of these mediator fields, e.g. a complex extension of $S$ charged under $U(1)'$ etc. (In the expression above, $B_{\mu\nu}$ is the SM $U(1)$ field strength, $L$ and $H$ are the SM Higgs and lepton doublets, 
$A$, $\lambda$, $Y_N$ and $\epsilon$ are so-called portal couplings, and $F'_{\mu\nu}$ is the field strength of the new $U(1)'$ group.) Once coupled to the SM through these channels, the IR hidden sector described by ${\cal L}_{\rm hid}$ can be almost arbitrarily complicated. $S$ and $N$ can couple to a complex hidden sector involving dark abelian or non-abelian gauge groups, possibly with additional scalar or fermion states charged under those hidden gauge groups. The full hidden sector Lagrangian simply needs to comply with the conditions above. The portal interactions in (\ref{portal}) are complete under the assumption that the SM is strictly neutral under the extra $U(1)'$. 

Dark sector models with states well below the electroweak scale are often best probed using high-intensity accelerator-based searches, and this program has been developed extensively over the past decade. Given the generic complementarity between direct accelerator probes of new physics, and the indirect reach of precision low energy observables, it is natural to explore the role of indirect searches for new physics in the context of dark sector scenarios. As a prime example, electric dipole moments (EDMs) of atoms, molecules and nucleons 
 have for many years provided important indirect constraints on $CP$-violating new physics at or above the electroweak scale \cite{Pospelov:2005pr}. In recent years, progress in the measurements of atomic and molecular EDMs in particular has been significant, with the ACME Collaboration recently pushing the limit on the electron EDM (in combination with an associated semileptonic operator) down to $1.1\times 10^{-29}e$cm \cite{ACME}. 
 
 The electron EDM occupies an important niche among flavour-diagonal $CP$-odd observables, as it cannot be efficiently generated 
 by the  $CP$-odd vacuum $\theta$-angle of QCD, the latter being  tightly constrained by the neutron EDM. 
 Instead, the current level of sensitivity to $d_e$ predominantly probes $CP$-violating physics all the way to 
 the $\sim100$\,TeV scale, constraining weak scale supersymmetry, 
 left-right models, multi-Higgs models, etc. Such models generically have in common the presence of extra 
 states charged under the SM gauge groups, which are in turn constrained to be heavy by collider bounds. The future discovery of a non-zero electron EDM would then appear to point to new physics at the weak scale or above. It is natural to try and test this conclusion in more detail, and explore whether there are mechanisms via which neutral dark sectors can induce EDMs, and $d_e$ in particular. This is the question that we will 
 address in the present paper.
 The existence of efficient dark sector mechanisms to generate $d_e$ close to the existing limits would weaken the direct connection 
 to new UV physics, and broader classes of dark sub-electroweak scale physics would then become relevant in interpreting an EDM discovery.
 
 The impact on EDMs of adding dark sector degrees of freedom coupled through the vector and scalar portals was analyzed briefly from a general perspective in \cite{LPR15}. It was observed that the addition of the vector portal introduced a new mediation mechanism for $CP$-violation, with the primary generation of `dark EDM' operators, associated with a coupling to a light dark photon $A'_\mu$. In turn, these operators preferentially induce higher-dimensional $CP$-violating EDM radius (or Schiff moment) operators for SM fermions rather than EDMs directly, which entails further suppression. The present paper builds on this analysis and 
 considers generic CP-violating SM-neutral dark sectors at the weak scale and below. 

The only portal coupling in (\ref{portal}) that allows for explicit $CP$-violation in mediation is the neutrino Yukawa $Y_N$,
while $\epsilon$, $A$ and $\lambda$ are explicitly real and $CP$-conserving. The potential contribution to EDMs from $Y_N$, contributing at 2-loop order, has been studied for some time \cite{Ng,Archambault:2004td,LPR15}, particularly given the motivation for additional $CP$-odd phases from leptogenesis. However, the contributions to lepton EDMs are generically suppressed well below the level of current sensitivity, $|d_e| < 10^{-33} e{\rm cm}$ \cite{Archambault:2004td}, in part by the small neutrino mass scale.\footnote{Note that relaxing the connection to neutrino mass generation via the neutrino portal allows for somewhat larger EDM contributions \cite{Abada:2015trh}.} 
The important point for a more general analysis is that ${\cal L}_{\rm hid}$ can provide {\em large} sources of $CP$-violation. 

 In the present work, we revisit the generation of EDMs from dark sectors. We clarify the leading (albeit suppressed) contribution to the electron EDM from light dark sectors, in the case that the electroweak scale and associated degrees of freedom are decoupled and thus only the vector portal survives. Furthermore, on allowing the portal interactions tied to the electroweak sector, we  identify a new combined mediation channel, which we term the {\it singlet portal}, involving both the neutrino and scalar portals, which can induce a sizeable electron EDM. The singlet portal allows EDM contributions which avoid the primary suppression factors noted above, and may in turn be naturally linked to models of baryogenesis. We will make the assumption that $CP$ phases in the hidden sector are maximal, and determine the scale of the induced EDMs, given the restrictions already in place on the portal couplings from a variety of other experimental probes. 
 
The rest of this paper is organized as follows. In Section~2, we analyse the EDMs from light dark sectors, discussing several generic mechanisms for transmitting $CP$-violation from the dark sector to electrons, including an important case when the weak scale is decoupled. Then in Section~3, we 
 generalize this concept to new physics at the electroweak scale, and focus on a new 2-loop EDM contribution mediated by the neutrino and scalar (singlet) portal. We calculate the 2-loop contribution to $d_e$ resulting from Higgs exchange, and examine the complementary sensitivity  that the ACME electron EDM constraint provides to the portal couplings, as compared to direct probes at the intensity and energy frontiers. The resulting sensitivity plots are shown in Fig.~4. We conclude in Section~4 with a discussion of related EDM contributions, and potential implications for models of baryo/leptogenesis.

\section{Portals and EDMs}

In this section, while not attempting an exhaustive classification, we will explore several potential contributions to EDMs from a generic dark sector, coupled to the SM only via the three portal interactions. We will focus on the EDM of electron, 
\begin{equation}
\label{definition}
{\cal L} = -\frac{i}{2} d_e \bar e\sigma^{\mu\nu} \gamma_5 e F_{\mu\nu}, 
\end{equation}
and a semileptonic $CP$-odd interaction, $C_S \times 2^{-1/2}G_F(\bar e i \gamma_5 e)(\bar NN) $.
We start our analysis by considering dark sector models residing well below the weak scale. 

\subsection{EDMs in the $\La_{\rm dark } \ll m_W$ limit}

In a generic model with new physics at or above the weak scale, it is often the case that the corrections to EDMs and corrections 
to anomalous magnetic moments $\mu$ can be of the same order of magnitude. Indeed, taking models of supersymmetry for example, {\em if} the $CP$-odd phases
are maximal, then it is natural to expect that the contribution of superpartners to both dipole types
is comparable, $$|d_e^{\rm max}({\rm SUSY})| \sim O(| \mu_e({\rm SUSY})| ).$$ 
This conclusion changes drastically 
if we consider a UV-complete dark sector at the scale $\Lambda_{\rm dark } \ll M_W$,
\begin{equation}
\label{ineq}
|d_e^{\rm max}({\rm dark~sector})| \ll  | \mu_e({\rm dark~sector})| .
\end{equation}
It is important to understand the origin of such an inequality. 

Consider, for example, a low-energy theory of photons and electrons, extended 
by light new states. As a toy example, let us take QED plus one new scalar $S$, coupled 
to electrons as follows,
\begin{equation}
\label{naive}
{\cal L}_{\rm int} = S \bar e ( y + i \tilde y \gamma_5) e. 
\end{equation}
This theory would induce corrections to the electron magnetic moment $\Delta \mu_e$ 
proportional to $y^2$ and/or $\tilde y^2$ (in units of $m_e/m_S^2$).  
When both couplings are present, an EDM also arises, $d_e \propto  y\tilde y$,
and no hierarchy of the form (\ref{ineq}) is apparent. In practice, the
relation (\ref{ineq}) arises due to the embedding of (\ref{naive}) into the full SM electroweak structure.
At tree level, the interaction (\ref{naive}) may only result from integrating out the SM 
Higgs field, via a combination of the $y_e LHE$ gauge invariant term in the 
SM Lagrangian and the $ASH^\dagger H$ portal interaction. In this case, it is a simple exercise to see that, {\it (i)} 
the Yukawa coupling $y$ does not exceed the electron Yukawa coupling, $|y| < |y_e|$, 
and {\it (ii)} the $CP$-odd Yukawa coupling is simply absent, $\tilde y = 0$. In order to generate a non-zero $\tilde y$
at tree level one needs either extra sets of Higgs doublets and/or extra massive vector-like fermions 
mixing with electron fields. All these types of interactions fall outside of our definition of dark sectors, as they contain charged particles. 

We conclude that the scalar portal is intrinsically $CP$-even after accounting for its SM embedding, and decouples in the formal limit $m_W \rightarrow \infty$. 
The neutrino portal also decouples in the same limit.
 Thus, if we wish to focus on physics that decouples from the electroweak scale, we need only retain the vector portals.
 These portals are $CP$-even as well, but can mediate low-scale $CP$-violation present in ${\cal L}_{\rm hid}$. 

Consider a dark $U(1)$ sector - possibly with mutliple gauge groups - interacting with the SM and containing $CP$-odd phases in the 
dark sector,
\begin{equation}
{\cal L}_{\rm int}= eJ^\mu_{\rm EM}\sum_a \epsilon_aA'^{a}_\mu +  {\cal L}_{\rm hid}(A'_a, S, H_d, \psi_d,...).
\end{equation}
Here $H_d, \psi_d$ denote scalar and fermionic fields charged under the dark $U(1)'$ gauge groups. 
$CP$-violation in ${\cal L}_{\rm hid}$ can be realized in a variety of ways, {\em e.g.} via 
Yukawa couplings: $\bar\psi_d (m_\psi + S(Y_S + i\tilde{Y}_S \gamma_5))\psi_d$. It is well-known 
that at loop level such interactions can generate effective $CP$-odd operators composed solely from 
$A'_a$ fields, with a minimum of two loops. We choose to classify such interactions 
in the soft limit, when the momentum of $A'_a$ is smaller than the loop momentum of the dark sector fields. 
In the final answer for $d_e$, this limit will not provide the dominant contribution. However, it allows us to correctly 
count the minimum number of loops and the order in coupling constants at which a nonzero answer may first appear. 

For a single dark $U(1)$ group, the minimal $CP$-odd operator that can be generated is of dimension 8,  
\begin{equation}
{\cal L}_{\rm hid} = \frac{c^{(2)}}{\La^4} (F'_{\mu\nu}F'^{\mu\nu})(F'_{\alpha \beta} \tilde F'^{\alpha \beta} ) +\cdots,
\label{4F}
\end{equation}
where $c^{(2)}$ is a coefficient reflecting the two-loop nature of the operator, and $ \La$ is the mass scale associated 
with the dark sector. An electromagnetic analog of the operator (\ref{4F}) would induce an electron 
EDM requiring two further loops with one of the 
field strengths being part of the EDM operator. In contrast, all dark gauge bosons need to integrated out requiring a loop diagram such as Fig.~1(top), 
with the electric field being attached to the electron line. As a result, we conclude that $d_e$ is first generated with an additional 
three loops, and at quartic order in the kinetic mixing parameter $\epsilon$. Symbolically, 
\begin{equation}
d_e^{\rm max} \sim \frac{m_e}{\La^2} \times \epsilon^4 \times c^{(2+3)}.
\end{equation}
Here we assume that the UV divergence generated by the operator (\ref{4F}) is stabilized at the scale $\Lambda$, 
and $c^{(2+3)}$ is the loop coefficient that accounts for two dark sector loops and three additional mediator loops as in Fig.~1(top). 
The resulting estimates are well below current experimental capabilities for the whole range of $\Lambda$ down to the GeV scale for example.

It is natural to ask if the existence of multiple $U(1)'$ groups might reduce the loop level for communicating $CP$-violation from ${\cal L}_{\rm hid}$
to $d_e$? The dark sector theory may generate three-boson dim=6 
operators such as $O_1 = F^a_{\mu\nu}F^b_{\nu\alpha}F^c_{\alpha\mu}$ and 
$O_2 = F^a_{\mu\nu}F^b_{\nu\alpha}\tilde F^c_{\alpha\mu}$,
provided that three or more distinct $U(1)'$ gauge groups are present. However, such operators 
do not lead to an EDM. To see this, let us assign to all dark $U(1)'$'s the same properties under all discrete symmetries as for the EM field $A_\mu$, {\i.e.} under charge conjugation, $C(A'^a)= - A'^a$ as for regular electromagnetism. This way the kinetic mixing portals respect all symmetries. Then, operator $O_1$ has 
$(-1,1,-1)$ transformation properties under $C,P,T$, while $O_2$ transforms as $(-1,-1,1)$. Notice that the transformation 
properties of an EDM  by (\ref{definition}) are $(1,-1,-1)$. Therefore, since the kinetic mixing and QED vertices conserve all discrete symmetries, 
neither $O_1$ nor $O_2$ lead to an EDM. Once mixing with the $Z$-boson is accounted for, which can invert the $C$ and $P$ parity because of 
its interaction with the axial-vector current, the operator $O_1$ can induce an EDM. 
(This is unlike the nonabelian case of the Weinberg operator \cite{Weinberg:1989dx}, 
where $CP$-violation requires the presence of the dual field strength, the trilinear CP-odd operator composed of distinct $U(1)$'s 
is $O_1$, as can be seen by reducing the field strengths to their electric and magnetic components.)
The mixing with the $Z$ is, however, suppressed by $G_F$.

Therefore, our analysis in the decoupling limit of weak interactions confirms the expectation that the {\em maximum} value of $d_e$
induced by the dark sector is always much smaller than its magnetic counterpart. The first non-vanishing correction to $d_e$ appears at five loops and is quartic in the kinetic mixing parameter. Corrections to $\mu_e$ can in contrast be generated
by a one-loop dark photon exchange diagram \cite{Fayet:2007ua,Pospelov:2008zw}.

As a stand-alone remark, we would like to add that dark sectors, realized in form of operator $O_1$ may provide 
an interesting test case for studying $T$-odd, $P$-even interactions. The usual problem is that such interactions 
easily transmute to those that are $T$-odd, $P$-odd, due to the fact that weak interactions easily 
flip parity, and therefore are tightly constrained by EDMs \cite{Khriplovich:1990ef}. However, $T$-odd, $P$-even interactions
that arise from dark sectors with $\Lambda \ll m_W$ may help to  circumvent this constraint.

\subsection{Dark Barr-Zee mechanism}

Lifting the $\La_{\rm dark} \ll M_W$ restriction and allowing for weak interactions opens up other possible portals. 
The presence of two or more different portal couplings leads to a number of alternate channels for generating an EDM at two-loop order. However, there are generically three portal mixing insertions required, which again implies significant suppression. As noted in \cite{LPR15}, adding the scalar to the vector portal allows the generation at 2-loop order of the `dark EDM' of a SM fermion such as the electron, 
\be
 d^{\rm dark}_e \bar{e} \si^{\mu\nu} i \gamma_5 e F'_{\mu\nu}.
 \ee
Importantly, since $A'_\mu$ is generically massive, integrating it out in the latter case generates a higher-dimension `EDM radius' (or Schiff moment), rather than an EDM,
\be
  \bar{e} \sigma^{\mu\nu} i\gamma_5 e F'_{\mu\nu}  \longrightarrow \frac{1}{m_{A'}^2}\bar{e} \sigma^{\mu\nu} i\gamma_5 e\, \Box F_{\mu\nu}.
  \ee
This is a characteristic feature of a light hidden sector with a kinetically mixed $U(1)'$ gauge group, namely that the naive dimension counting of operators in the SM effective field theory can be misleading as high-dimensional operators may be numerically enhanced by powers of ($m_W/\La_{\rm dark}$) and require less loop factors. Nonetheless, the resulting observable EDM is still relatively suppressed, particularly given the current limits on the kinetic mixing parameter $\ep$. 

As an explicit example, a generic Barr-Zee-type contribution is shown in Fig.~1~(middle) \cite{LPR15}. This involves the $CP$-odd coupling of the scalar $S$ to a dark sector fermion $\ps$, $\bar\psi (m_\psi + S(Y_S + i\tilde{Y}_S \gamma_5))\ps$, which is in turn charged under $A'_\mu$ \cite{Bird:2006jd,O'Connell:2006wi,Barger:2007im,Sato:2011gp,Fox:2011qc,Low:2011gn}.  Integrating out $\ps$ leads to a $CP$-odd $SF'\tilde{F}'$ operator, which in turn can generate the dark EDM of an electron. As discussed above, this will generate the electron EDM radius rather than the SM EDM directly. The contributing diagrams were analyzed in \cite{LPR15} and, with the hierarchy $m_{A'} \ll m_S \ll m_\ps$ and Higgs-scalar mixing angle $\theta_h \ll 1$, one finds the EDM radius,
\be
   r^2_{d_e}  = \frac{|e| \alpha' \tilde{Y}_S m_e}{16\pi^3 v m_\ps m_{A'}^2} \times  \ep^2 \theta_h  \ln(m_\ps^2/m_S^2).
\ee
An estimate of the resulting EDM probed in atomic/molecular EDM experiments, induced by the mixing of $s-p$ orbitals, follows by identifying the corresponding scale with the $K$-shell radius, $d_e \sim (Z \alpha m_e)^2  r^2_{d_e}$, given $m_{A'} > Z \alpha m_e$. This leads to the estimate \cite{LPR15}
\be
 d_e \sim \left(4\cdot 10^{-33}~e\cdot {\rm cm}\right) \times \left(\frac{1\,{\rm GeV}}{m_\psi}\right)\left(\frac{\ep}{10^{-4}}\right)^2 \left( \frac{\theta_h}{10^{-3}}\right),
\ee
which is still well below the current sensitivity to the electron EDM, due in part to the strong limit on $\ep$ in the relevant $A'_\mu$ mass range from $g-2$ of the electron \cite{Pospelov:2008zw} and direct searches for dark photon at NA64 \cite{Banerjee:2017hhz} and BaBar \cite{Lees:2017lec}. Note that the $SFF'$ operator can also generate the semi-leptonic interaction $C_S \bar{N}N \bar{e} i \gamma_5 e$, which is also probed in paramagnetic EDM experiments. However, this again requires two loops and three mixing insertions.

\begin{figure}[t!]
\centerline{\includegraphics[viewport=80 420 600 820, clip=true, scale=0.33]{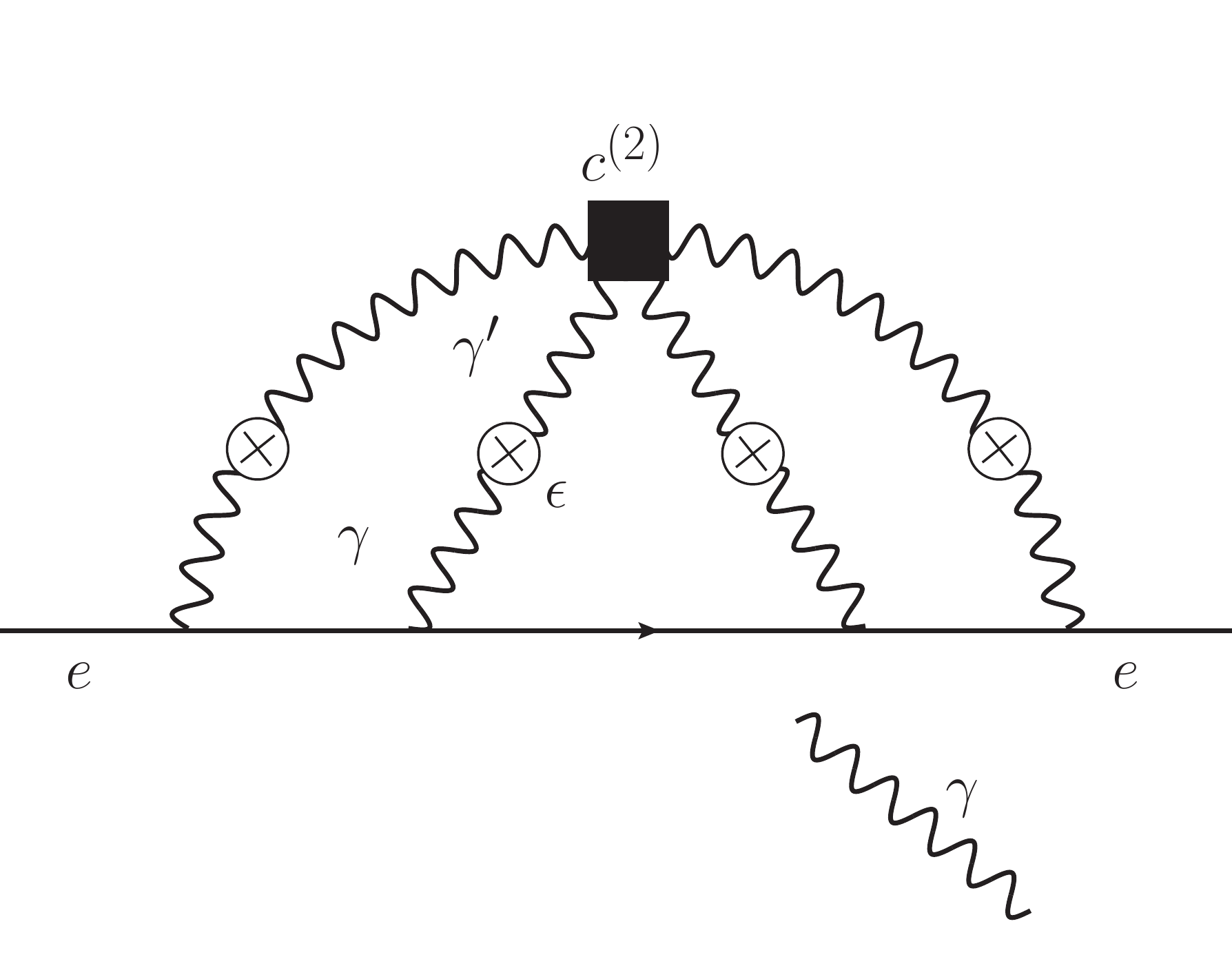}}
\centerline{\includegraphics[viewport=140 420 440 720, clip=true, scale=0.35]{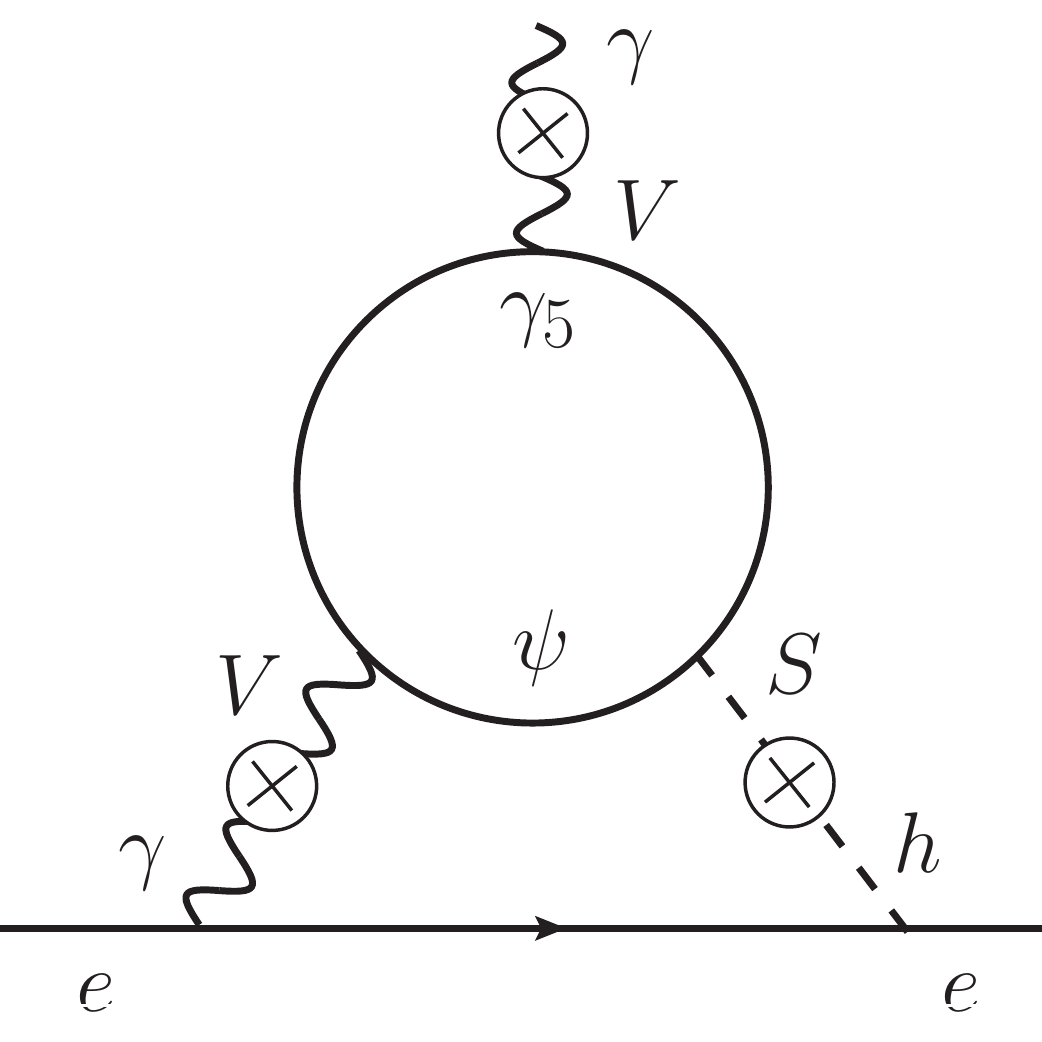}}
\centerline{\includegraphics[width=6.5cm]{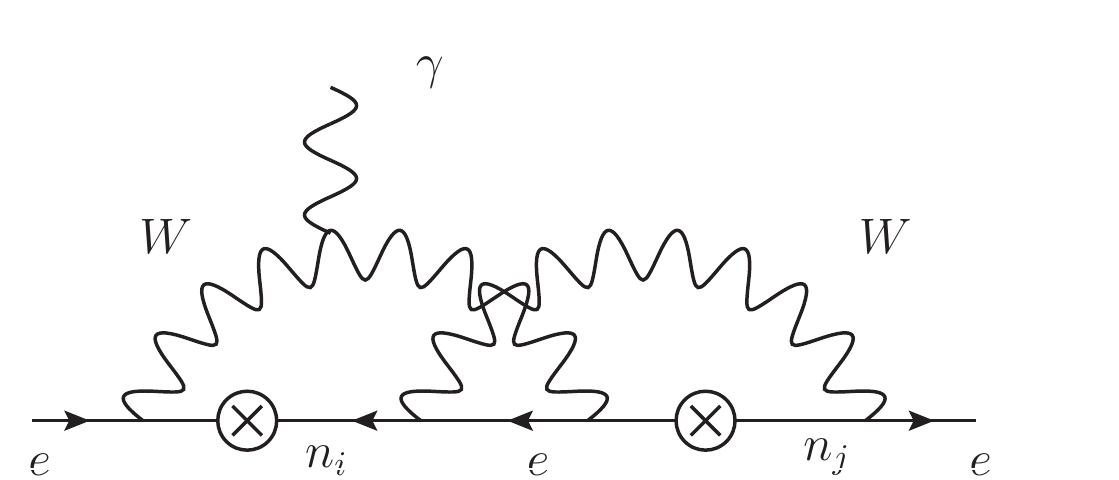}} \vspace*{0.5cm}
\caption{Three suppressed hidden sector contributions to lepton EDMs: (top) the leading 3-loop contribution due to a dark $U(1)'$ and the operator (\ref{4F}); (middle) a 2-loop contribution induced by the dark EDM of the electron, mediated by the vector and scalar portals; and (bottom) an example of the leading 2-loop contribution due to $CP$-violation in the neutrino portal. The crosses denote portal couplings. See the text for further details.} 
\label{fig:edm}
\end{figure}

Turning on all three portals does not appear to quantitatively change the characteristic EDM contributions which still require two loops and three mixing insertions, but does introduce some additional freedom in the choice of the insertions. Rather than exploring these contributions more systematically, in the next section we will focus on a different channel which appears to provide the largest EDM contribution, given current constraints on the portal mixing angles.

\section{EDMs from the singlet portal}

The neutrino portal $LHN$ allows for $CP$-violation in the portal interaction $Y_N$ itself.
As the simplest seesaw model for neutrino mass generation, and for leptogenesis, the neutrino portal has been studied extensively, including its contribution to lepton EDMs. For completeness, we summarize the conclusions here. In order to incorporate a nontrivial $CP$ phase, we require two singlet fermions $N_R, N_S$ leading to the following mass matrix for $(\nu_L, N_R,N_S)$,
\be
   {\cal M} = \begin{pmatrix}
                    0 && m_{D_1} && m_{D_2}\\
                    m_{D_1} && M_R && \De M\\
                    m_{D_2} && \De M && M_S
                   \end{pmatrix},
\eeq
where $m_{D_i}$ are the Dirac masses and we work in the regime $M_{R,S}\gg m_{D_i},\De M$, where the mass spectrum includes a light neutrino $m_\nu \simeq(m_{D_1}^2-m_{D_2}^2)/M$, and heavy states $M_+^0\simeq M_S$, $M_-^0\simeq M_R$, 
with $M=(M_R+M_S)/2$ the Majorana mass scale. As shown in Fig.~1(bottom), an EDM is generated at two loop order \cite{Archambault:2004td,LPR15},
\beq
\begin{split}
 d_e\sim&\left(3\cdot10^{-35}~ e\,\text{cm}\right)\frac{m_{D_1}^2m_{D_2}^2}{M^4}\frac{M_S^2-M_R^2}{\text{GeV}^2}.
\end{split}
\eeq
The ratios $\theta_\nu \sim m_{D_i}/M\lesssim10^{-1}$ are the visible-hidden mixing angles and, even with considerable tuning, the constraints on the light neutrino mass spectrum limit the EDM to less than $10^{-33}\text{e}\cdot\text{cm}$, well below the current experimental limit. As noted earlier, dropping the connection of the neutrino portal to neutrino mass generation, and admitting more general mixing, allows for somewhat larger EDM contributions \cite{Abada:2015trh}.

\begin{figure}[t!]
\centerline{\includegraphics[viewport=100 420 620 750, clip=true, scale=0.3]{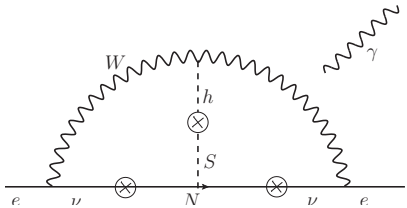}}
\caption{A 2-loop contribution to lepton EDMs mediated by the combined singlet portal.} 
\label{fig:edm}
\end{figure}

If we combine the neutrino and singlet scalar portals, 
\be
 {\cal L}_{\rm singlet} = - AS H^\dagger H - (Y_N LHN +(h.c.)),
\ee
a combination that we term the {\it singlet portal}, a new 2-loop contribution to EDMs can be identified as shown in Fig.~2. This channel does not require any additional dark sector degrees of freedom, or multiple generations of fermions. The addition of the scalar portal allows a number of the suppression factors impacting Fig.~1(bottom) to be avoided, and this diagram is parametrically quite large given the limited constraints on the neutrino and scalar mixing angles. Most importantly, $N$ can be chosen to be a Dirac particle, 
and thus unconstrained by the visible neutrino mass splitting. 

In this section, we will calculate this EDM contribution explicitly, and explore the complementarity of the EDM sensitivity to other direct collider and fixed target probes. 
The calculation can be performed at leading order in the weak interactions ($\alpha_W \to 0$), 
utilizing only the Goldstone components $G^\pm$ of the weak vector bosons.  After electroweak symmetry breaking, the relevant portal couplings take the form,
\begin{align}
{\cal L}_{\rm singlet} &= \frac{\theta_h}{v} (m_S^2 - m_h^2) \, S \left( vh + G^+G^- \right) + \lambda_N S \bar N\gamma_5 N \nonumber\\
&+ \frac{\sqrt{2}\theta_\nu}{v} \left[ G^-\bar e \left( m_e P_L  + m_N P_R\right) N +h.c. \right]+\cdots 
\end{align}
where $\theta_h = Av/(m_S^2-m_h^2)$ is the scalar mixing angle, which tends to $\theta_h \rightarrow Av/m_h^2$ for $m_S \ll m_W$, while $\theta_\nu$ is the corresponding singlet neutrino mixing angle. We have introduced a further pseudoscalar hidden sector coupling $\lambda_N$ between $S$ and $N$, which therefore breaks $CP$ in the full  theory and allows for EDMs at 2-loop order. Note that on diagonalizing the mass matrices, this will also induce a $\nu N S$ coupling proportional to $\lambda_N \theta_\nu$. From the structure of the diagrams in Fig.~2, the  characteristic length scale $L_e^{\rm scale}$ of the 2-loop contribution to $d_e$ takes the form,
\begin{align}
L_e^{\rm scale} &= \frac{\lambda_N \theta_\nu^2 \theta_h}{(16\pi^2)^2} \times \frac{2m_e m_N}{v^3} \nonumber\\
 &\sim 4 \times 10^{-27}\,{\rm cm} \times \lambda_N \theta_\nu^2 \theta_h \times \frac{m_N}{m_W}. \label{dscale}
\end{align}

It is convenient to extract the EDM from the more general 2-loop amplitude for the electron self-energy in a general electromagnetic background field $F_{\mu\nu}$. This background is incorporated by retaining the dependence on the covariant electron momentum $P_\mu = p_\mu + e A_\mu$, which satisfies $\slash{P} e(P) = m_e e(P)$ and $[P_\mu, P_\nu] = i e F_{\mu\nu}$. In terms of the overall length scale ({\ref{dscale}), and the loop momenta $q$ and $k$, the corresponding amplitude can be schematically written as follows,
\begin{align}
{\cal M} = e L_e^{\rm scale} \times \int d^4qd^4k  \bar e(P) [ f_{\rm F}(q,k)\times  f_{\rm B}(q,k,P)  ] e(P) 
\label{eq:amp}
\end{align}
where $f_{\rm F}$ and $f_{\rm B}$ are respectively fermionic and bosonic integrands, that are separated to emphasize that the electric charge flows over the charged Goldstone line, and therefore only the bosonic integrand depends on the covariant electron momentum $P_\mu$. The amplitude is to be understood as a Taylor series in $P_\mu$ and $m_e$.

\begin{figure}[t]
\centering
 \includegraphics[width= 0.48\textwidth]{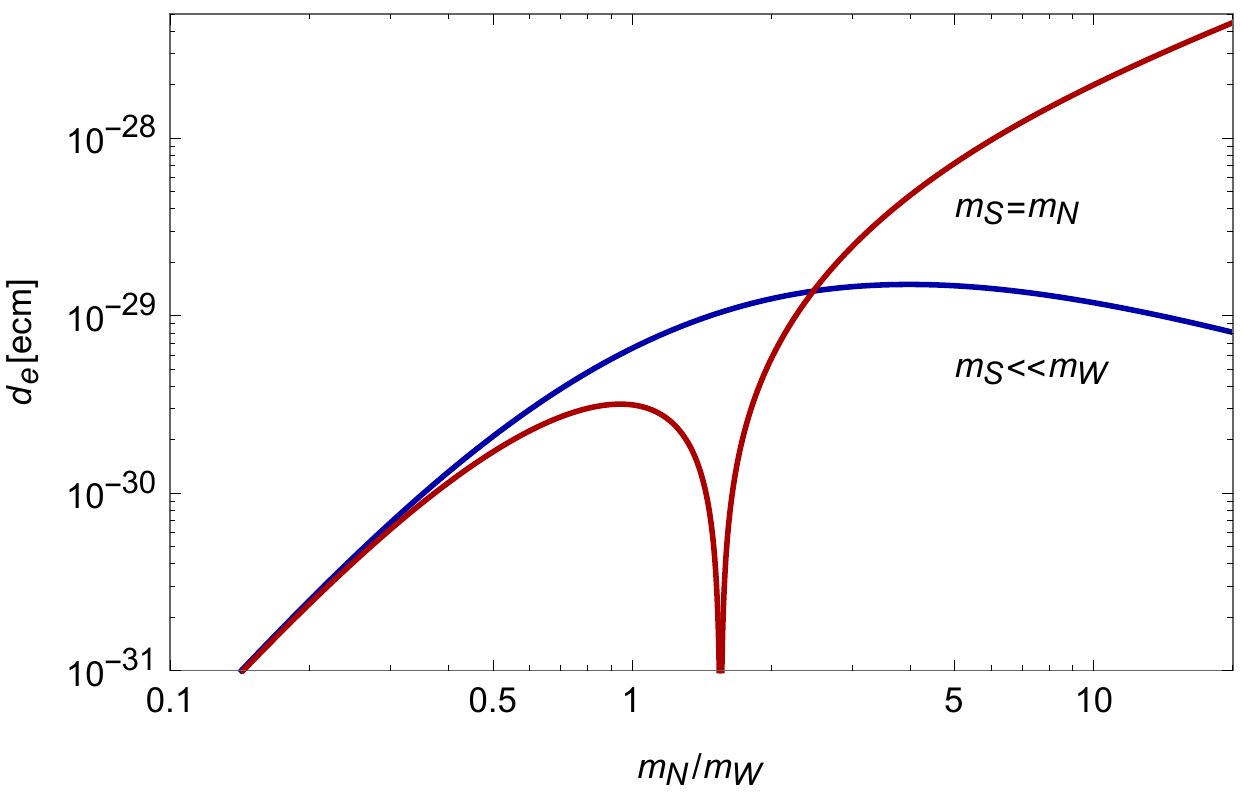}
  \caption{ For the  singlet $N-S$ portal example, the size of the electron EDM is shown as a function of $m_N/m_W$ for two different choices of the singlet scalar mass: $m_S \ll m_W$ and $m_S=m_N$. We have set $\theta_h \theta_\nu^2 = 10^{-2}$.}
\label{fig:edms}
\end{figure}

Further details of the calculation are outlined in Appendix A. Here we simply note that after expanding the integrands up to third order in $P_\mu$ and computing the relevant commutators, one can isolate all EDM contributions in the form
\begin{align}
   {\cal M} = - \frac{i}{2} eL_e^{\rm scale} \bar{e} \sigma^{\mu\nu}\gamma_5 e F_{\mu\nu}\times \int d^4qd^4k f_{\rm scalar} (q,k),
   \label{eq:fscalar}
\end{align}
so that 
\be
 d_e = e L_e^{\rm scale} \times \int d^4qd^4k f_{\rm scalar} (q,k),
\ee
where $f_{\rm scalar}$ is a dimensionless scalar integrand. After integration over $q$, the results can be expressed in terms of Passarino-Veltman functions, which are conveniently handled using {\tt Package-X} \cite{Patel:2015tea}. The resulting dimensionless function $\int d^4qd^4k f_{\rm scalar} (q,k)$ is suppressed for $m_N$ well below the weak scale. Numerical results in two scaling regimes for $m_S$ are shown in Fig.~3. The resonant behaviour for $m_S\rightarrow m_h$ is apparent in the case that $m_S = m_N$, with the suppression of the EDM due to the fact that $\theta_h = Av/(m_S^2-m_h^2)$ is held fixed. The relatively weak decoupling as $m_N \rightarrow \infty$ is reminiscent of the dependence of many FCNC observables on $m_t$. The full loop function has a complex dependence on the mass scales, but the following scaling limits discussed further in Appendix B are more illuminating. 

In the limit where $m_W \ll m_S, m_N$, we find
\begin{align}
 \int d^4qd^4k  f_{\rm scalar} (q,k) \rightarrow \left\{ \begin{array}{ll}  \frac{3}{4} \ln\frac{m_S^2}{m_N^2}, & m_N \ll m_S \\ 
    \frac{1}{4}\frac{m_S^2}{m_N^2} \ln \frac{m_N^2}{m_S^2}, & m_S \ll m_N \end{array} \right. . \label{scaling0}
\end{align}
Alternatively, if we keep $m_N$ finite and scale $m_S/m_W$, 
\begin{align}
 \int d^4qd^4k  f_{\rm scalar} (q,k) \rightarrow \left\{ \begin{array}{ll}  \frac{3}{4}\frac{m_N^2}{m_W^2} \ln\frac{m_S^2}{m_W^2}, & \frac{m_S}{m_W} \rightarrow \infty \\ 0.95\frac{m_N^2}{m_W^2}, & \frac{m_S}{m_W} \rightarrow 0 \end{array} \right. , \label{scaling1}
\end{align}
while for $m_N\rightarrow \infty$,
\begin{align}
  \int d^4qd^4k  f_{\rm scalar} (q,k)  \rightarrow \left\{ \begin{array}{ll}  0.54, & m_S = m_N \\ \frac{1}{4}\frac{m_h^2}{m_N^2} \ln\frac{m_N^2}{m_W^2}, & \frac{m_S}{m_W} \rightarrow 0 \end{array} \right. . \label{scaling2}
\end{align}
Note that in both cases, the decoupling in the limit $m_S \rightarrow \infty$ is not visible in the loop functions, but arises from $\theta_h \propto 1/m_S^2$ with our normalization of the $hS$ mixing vertex.

The logarithmic mass dependence in (\ref{scaling0})  can be understood as the result of renormalization group (RG) running between $m_S$ and $m_N$ (or vice versa) in an effective field theory analysis. For example, if we focus on the regime $v \ll m_N \ll m_S$, the singlet scalar can be integrated out at the high scale $m_S$, leading to an effective $CP$-odd operator, $\bar{N} i \gamma_5 N (H^\dagger H)$. This operator generates an EDM, with a log-divergent loop and a coefficient of 3/4, precisely matching the scaling limit in (\ref{scaling0}). A similar argument can be used to interpret the logarithmic scaling in (\ref{scaling1}) and (\ref{scaling2}).

To exhibit a characteristic scale for the EDM,  in the mass range $m_N \sim (1-10)\, m_W$, with $m_S \ll m_W$, we find numerically
\be
 d_e  \sim 10^{-29}e\,{\rm cm} \times \left(\frac{\lambda_N \theta_\nu^2 \theta_h}{10^{-2}}\right) .
\ee
This is quite close to the current experimental limit from ACME of $1.1\times 10^{-29}e$cm, assuming relatively mild constraints on the mixing angles, and is the largest EDM contribution we have uncovered according to our definition of a UV-complete dark sector. The full scaling of the EDM with $m_N$ and $m_S$ is illustrated in Fig.~4 for two different portal mass regimes, where we also exhibit the leading direct constraints on the mixing angles for comparison. The constraints on the neutrino mixing angle are from a CHARM search for heavy neutral leptons \cite{CHARM}, DELPHI searches for $Z\rightarrow N\nu$ \cite{DELPHI}, an ALEPH measurement of $W$-pair production \cite{ALEPH,AF15}, and precision electroweak data (EWPD) including lepton universality \cite{AF15}. The relevant limits in each mass range are combined with limits on Higgs-scalar mixing from an L3 search for $e^+e^-\rightarrow Z^* S$ \cite{L3}, to produce the limit contours in Fig.~4. For $m_S \ll m_W$, the limit on Higgs mixing is $\theta_h \sim 0.1$ \cite{L3}, while it weakens to ${\cal O}(1)$ for larger $m_S$. This results in the slightly weaker direct limits shown in Fig 4 (bottom) for larger masses. Note that the stringent limits on lepton flavour violating processes cannot be directly applied to $\theta_\nu$ without further assumptions about the flavour structure of the $N$-sector. 

\begin{figure}[t]
\centering
 \includegraphics[width= 0.48\textwidth]{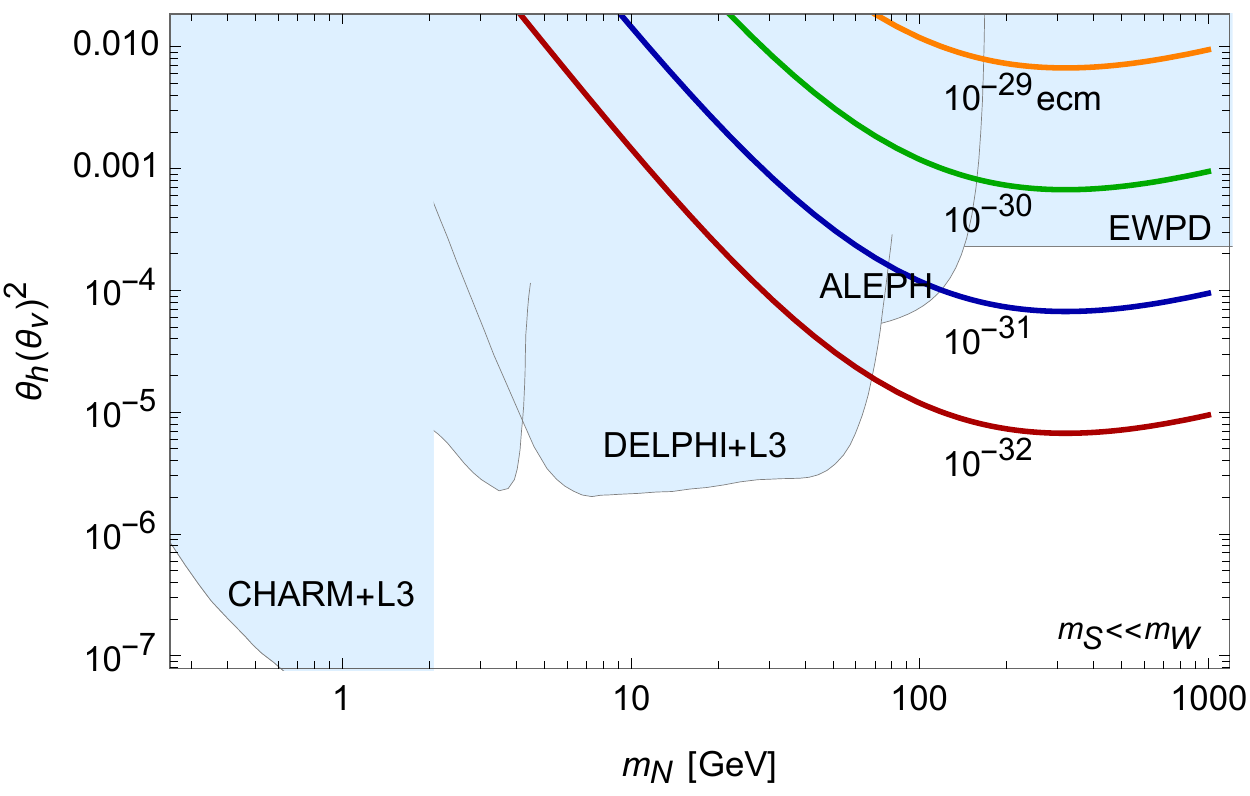}
 \includegraphics[width= 0.48\textwidth]{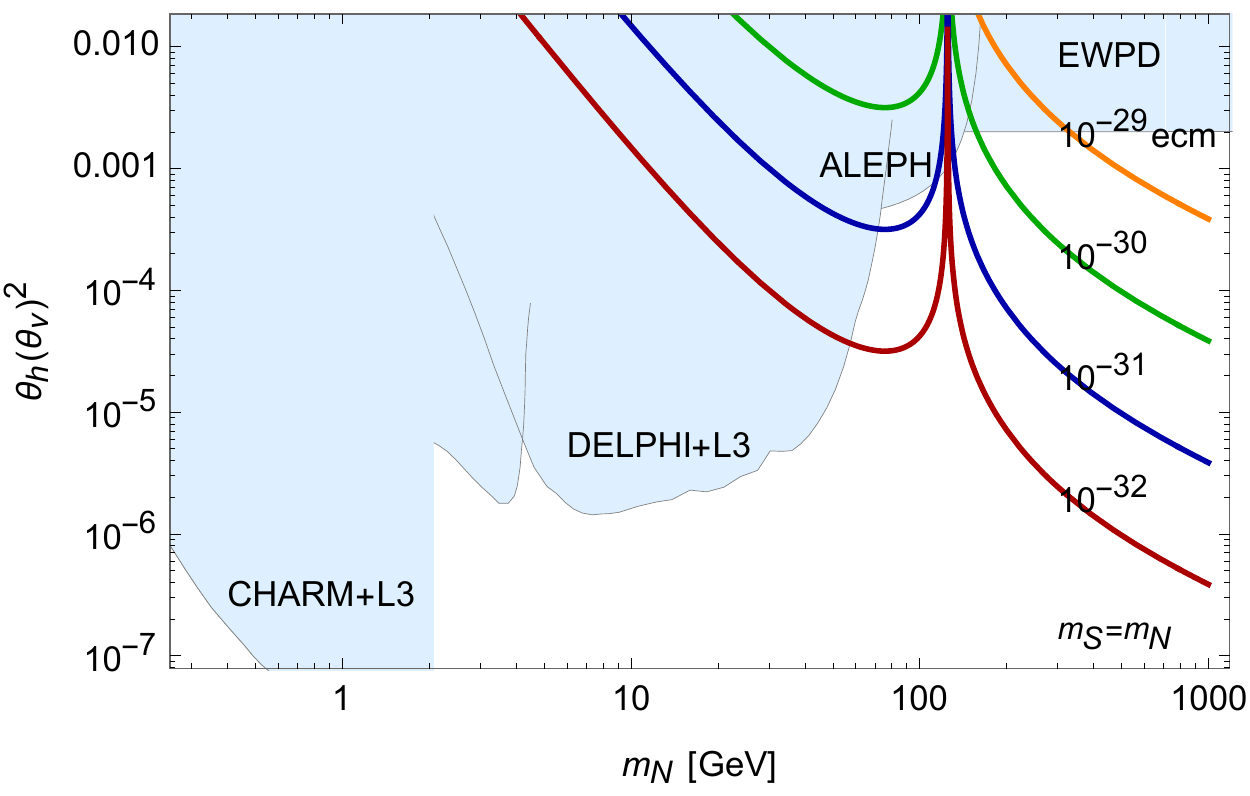}
  \caption{ For the  singlet $N-S$ portal example, EDM sensitivity contours (with a maximal $CP$ phase) are shown as a function of the singlet neutrino mass, assuming (top) $m_S \ll m_W$ and (bottom) $m_S = m_N$. Note that the current limit from the ACME collaboration is $1.1\times 10^{-29}e$cm. For comparison, the limits on the portal couplings are shown in light blue. The limits on $\theta_h$ are from L3 \cite{L3}, while the limits on $\theta_\nu$ are from a range of measurements as shown \cite{CHARM,DELPHI,ALEPH,AF15}. See the text for further details. }
\label{fig:limits}
\end{figure}

\section{Conclusions}

The majority of the EDM contributions from dark sector $CP$-violation are below current experimental sensitivity, primarily due to the significant constraints on the portal couplings from independent measurements. In particular, in the limit of light dark sectors, $\Lambda_{\rm dark} \ll m_W$,
the leading diagrams that can generate an electron EDM independent of weak sector degrees of freedom appear at five loop order, and require four powers of the kinetic mixing parameter $\epsilon$ (which is typically limited to be less than $\sim 10^{-3}$ when the mass of the corresponding vector mediator is within 
reach of {\em e.g.} $B$-factories).

Nonetheless, we have identified contributions from the singlet $N-S$ portal that can be sizeable, and it is apparent that EDMs can provide sensitivity to the portal mixing angles that is complementary for large $m_N$ due to the mild decoupling behaviour. Moreover, the current limit on the electron EDM from the ACME experiment already provides sensitivity to this model that is comparable in reach to collider probes.
Should the next round of improvements (by ACME and/or other collaborations searching for $d_e$) lead to a positive detection, 
one would not be able to unambiguously assign it to models of new physics with charged particles. It could also be  
a signature of neutral dark sectors near the weak scale. For models with electroweak scale singlet fermions $N$, the upcoming high-luminosity run 
of the LHC may provide the best probe. 

We have focused in this paper on contributions to the electron EDM, given the recent dramatic improvements in experimental limits, and the fact that the three renormalizable portals provide more channels to the leptonic sector than to the hadronic sector. However, it is important to note that observable atomic/molecular EDMs can also be generated by semileptonic operators such as $C_S \bar{e}i\gamma_5 e \bar{N}{N}$, and the hidden sector scenarios studied here would also generate these operators. We have also focused on hidden sector mass scales of a GeV and above, but the possibility of mediators with smaller masses, closer to nuclear or atomic binding energies, may preferentially induce operators that are non-local on those scales. It would be interesting to explore deviations from 
the contact limit in this lower mass regime (see {\em e.g.} \cite{Gharibnejad:2014kda,Stadnik:2017hpa}). 
However, once the relevant mass scale becomes too small, 
(100 keV and smaller), typically, very strong astrophysical bounds on portal couplings can be imposed, leaving little scope for interesting sensitivity via static EDM observables. We also note that for all parameters considered here, the dark sector degrees of freedom decay to the SM well before BBN and thus have no cosmological impact. However, consideration of much lighter mass scales may bring cosmological constraints into play.

In conclusion, it is worth recalling that the primary empirical motivation for new sources of $CP$ violation is the need for a viable mechanism of baryogenesis. Indeed, the basic paradigm of leptogenesis, with $CP$ phases originating in the singlet (or right-handed) neutrino sector, falls into the category of dark sector $CP$-violation. The model considered here operates with mostly Dirac heavy neutrinos, that share the lepton number with SM leptons, pointing to electroweak baryogenesis as a promising scenario for this model. 
The extension to incorporate the other renormalizable portals opens up further channels to mediate this symmetry violation to the SM.
The additional singlet scalar, as is well known, can be used to induce/enhance a first order electroweak phase transition. 
 Recent applications of dark sector $CP$-violation to mechanisms of baryogenesis include \cite{Dall:2014nma,Cline:2017qpe,Carena:2018cjh}.

\begin{acknowledgments}
The work of  S.O, M.P. and A.R. is supported 
in part by NSERC, Canada, and research at the Perimeter Institute is supported in part by the Government 
of Canada through NSERC and by the Province of Ontario through MEDT. 
\end{acknowledgments}

\appendix
\renewcommand{\theequation}{\thesection.\arabic{equation}}

\section{Singlet portal EDM calculation}

This Appendix provides some details of the calculation of the relevant 2-loop EDM contribution from Fig.~\ref{fig:edm}. The approach taken was to calculate the electron self-energy in a general electromagnetic background field, and then expand the $CP$-violating part of the self-energy function in terms of the electron covariant momentum, $P_\mu = p_\mu + e A_\mu$. In the expansion, we only need to retain terms of order $P^3$ or $m_e P^2$, which will provide the leading contribution to the EDM. For reference, the momentum assignment is shown in Fig.~\ref{fig:app}. 

As mentioned in Sec.~3, the corresponding amplitude can be written in the form of Eq.~(\ref{eq:amp}), with loop integrals over $q$ and $k$ of separated bosonic and fermionic integrands. The bosonic part consists of the Goldstone, Higgs and singlet scalar propagators and is explicitly given by 
\begin{align}
f_{\rm B} (k,q,P) &= \frac{1}{\pi^4} \left[ \frac{m_S^2-m_h^2}{k^2-m_S^2} + \frac{(m_S^2-m_h^2)m_h^2}{(k^2-m_S^2)(k^2-m_h^2)} \right] \times \nonumber\\
& \frac{1}{(q-k/2-P)^2-m_W^2} \frac{1}{(q+k/2-P)^2-m_W^2} .
\end{align}
This is the only part dependent on the covariant electron momentum $P_\mu$. The fermionic part originates from internal propagating neutrinos, for which there are two possible combinations: heavy-heavy ($N$-$N$) and light-heavy ($\nu$-$N$). Note that the bosonic integrand is independent of the neutrino combinations.

The fermionic integrands are
\begin{align}
\label{eq:fF_NN}
f_{\rm F}^{(NN)} (k,q) &= \frac{1}{(-)_N(+)_N} \left[ i \frac{m_N^2}{m_e} \slash{k} P_L  \right.\\
&\left. + \left(m_N^2 - q^2 + \frac{k^2}{4}\right) i \gamma_5 + \sigma_{\alpha\beta} k^\alpha q^\beta \gamma_5 \right],
\nonumber
\end{align}
and
\begin{align}
\label{eq:fF_nuN}
f_{\rm F}^{(\nu N)} (k,q) &= -\frac{1}{2}\left[ \frac{1}{(-)(+)_N} + \frac{1}{(+)(-)_N}\right] \nonumber\\
& \times \left[  \left(q^2 + \frac{k^2}{4}\right) i \gamma_5 + \sigma_{\alpha\beta} k^\alpha q^\beta \gamma_5 \right],
\end{align}
where we use the shorthand notation, $(\pm)_N = (q \pm k/2)^2 - m_N^2$ and $(\pm) = (q \pm k/2)^2$, for heavy and light neutrino propagators respectively. To demonstrate the procedure, let us focus on the first line of Eq.~(\ref{eq:fF_NN}), for which the bosonic integrand should be expanded to order $P^3$, since the electron mass is included in the overall factor $L_e^{\rm scale}$. The fermionic integrand is odd in $k$, so it is convenient to antisymmetrize the bosonic integrand in $k \to-k$,
\begin{align}
f_{\rm B} (k,q,P) \to \frac{1}{2} \left( f_{\rm B} (k,q,P) - f_{\rm B} (-k,q,P) \right) ,
\label{eq:}
\end{align}
which is nonzero only when $P_\mu$ is noncommutative. Therefore, the expansion of Eq.~(\ref{eq:}) at order $P^3$ contains a commutator $[P_\mu, P_\nu]$ which in turn is replaced with the electromagnetic field strength $F_{\mu\nu}$. Once we obtain the terms containing $F_{\mu\nu}$, the remaining calculation can be simplified by replacing $P_\mu$ with the ordinary electron momentum $p_\mu$. Finally, making use of gamma matrix algebra and the equations of motion, the EDM operator can be factored out of the integrand before performing loop integrals. As a result, the amplitude can be written in terms of a Lorentz scalar integral over $k$ and $q$ as in Eq.~(\ref{eq:fscalar}).

The remaining contributions can be handled in a similar manner, utilizing $k \to -k$ (anti)symmetrization so that the amplitude expanded in powers of $P_\mu$ can always be expressed in terms of either $P^2$ or $[P_\mu, P_\nu]$, which in turn are traded for $F_{\mu\nu}$ via the relations $\slash{P}\slash{P} = P^2 + \frac{1}{2}eF_{\mu\nu}\sigma^{\mu\nu}$ and $[P_\mu, P_\nu]=ieF_{\mu\nu}$. The resulting EDM operator can be isolated from the integrand, leaving scalar integrals. After lengthy but straightforward calculations, we find the following contributions to the scalar integrand
\begin{align}
f_{\rm scalar}^{(NN)} &(q,k) = \frac{1}{\pi^4} \frac{k^2(m_S^2-m_h^2)}{(k^2-m_S^2)(k^2-m_h^2)} \times \nonumber\\
& \frac{1}{(+)_N(-)_N} \bigg\{ m_N^2 \left[ \frac{k^2/4}{(+)_W^2(-)_W^2} + \frac{2}{3} \frac{(kq)^2 - k^2q^2}{(+)_W^2(-)_W^3} \right] \nonumber\\
&+ (m_N^2 - q^2 + k^2/4) \left[ \frac{2m_W^2}{(+)_W^3(-)_W} + \frac{q^2 - k^2/4}{(+)_W^2(-)_W^2} \right] \nonumber\\
&+ \frac{1}{3} \frac{k^2q^2 - (kq)^2}{(+)_W^2(-)_W^2} \bigg\} ,
\end{align}
and
\begin{align}
f_{\rm scalar}^{(\nu N)} & (q,k) = \frac{1}{\pi^4} \frac{k^2(m_S^2-m_h^2)}{(k^2-m_S^2)(k^2-m_h^2)} \times \nonumber\\
&\left( \frac{-1/2}{(+)(-)_N} + \frac{-1/2}{(+)_N(-)} \right) \bigg\{ ( -q^2 + \frac{k^2}{4} ) \times \\
& \left[ \frac{2m_W^2}{(+)_W^3(-)_W} + \frac{q^2 - k^2/4}{(+)_W^2(-)_W^2} \right] + \frac{1}{3} \frac{k^2q^2 - (kq)^2}{(+)_W^2(-)_W^2} \bigg\} ,
\nonumber
\end{align}
where again $(\pm)_W = (q \pm k/2)^2 - m_W^2$, and $(\pm) = (q \pm k/2)^2$. Note that the two contributions precisely cancel in the limit $m_N \to 0$. The dimensionless scalar integrand introduced in Eq.~(\ref{eq:fscalar}) is just the summation of the above two functions. 

To carry out the first loop integral (over $q$), we employed {\tt Package-X}~\cite{Patel:2015tea} after which the result is expressed in terms of Passarino-Veltman functions. The remaining loop integral (over $k$) was performed numerically.

\begin{figure}[t]
\centering
 \includegraphics[viewport=100 610 480 770, clip=true, scale=0.5]{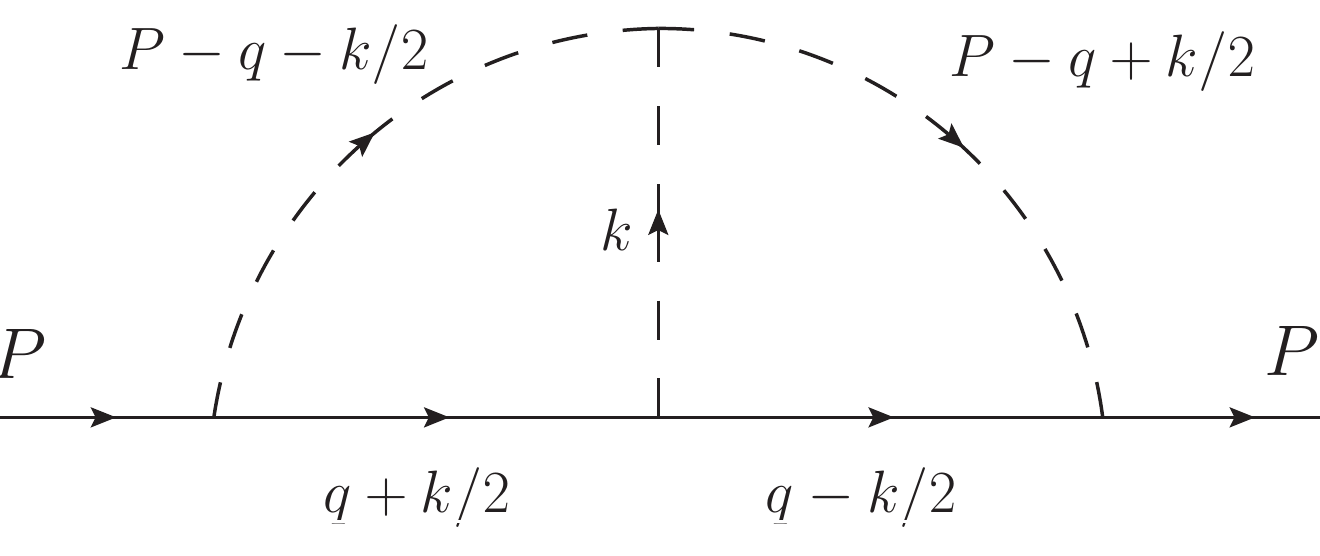}
  \caption{Momentum assignment in the relevant 2-loop diagram contributing to the electron self-energy, with arrows indicating the orientation.}
\label{fig:app}
\end{figure}

\section{Asymptotic scaling of the loop function}

This Appendix briefly summarizes the calculation of the scaling limits of the loop function shown in (\ref{scaling0}). Other limits can be obtained in a similar manner.

\subsection{$v \ll m_N \ll m_S$}
The loop function resulting from the integral over $q$ can be expanded in powers of the weak scale $v$,
\begin{align}
\int d^4q \, f_{\rm scalar} (q,k) \simeq g(k^2,m_N,m_S) + {\cal O}\left(\frac{v^2}{m_N^2}\right) ,
\end{align}
where the function $g(k^2,m_N,m_S)$ falls off faster than $1/k^4$ as $k^2\to\infty$. Let us now divide the integration region into two parts at an intermediate scale $\mu$ that satisfies $m_N \ll \mu \ll m_S$: $\int d^4k = (\int_{|k|\leq\mu} + \int_{|k|>\mu}) d^4k $. Expanding $g(k^2,m_N,m_S)$ in powers of $k^2/m_S^2$ for the former region, and in powers of $m_N^2/k^2$ for the latter region, and performing the integrals, we obtain
\begin{align}
\int d^4k \, g(k^2,m_N,m_S) &\simeq \frac{3}{4} \ln\left(\frac{m_S^2}{m_N^2}\right) - \frac{1}{4} ,
\end{align}
up to corrections of ${\cal O}(\frac{m_N^2}{m_S^2}, \frac{\mu^2}{m_S^2}, \frac{m_N^2}{\mu^2})$.

\subsection{$v \ll m_S \ll m_N$}
Following the same procedure as above, the $k$ integral is divided at $\mu$, where $m_S \ll \mu \ll m_N$, and we find
\begin{align}
\int d^4k \, g(k^2,m_N,m_S) &\simeq \frac{m_S^2}{m_N^2} \left[ \frac{1}{4} \ln\left(\frac{m_N^2}{m_S^2}\right) + \frac{1}{3} + \frac{\pi^2}{36} \right] ,
\end{align}
up to corrections of ${\cal O}(\frac{m_S^2}{m_N^2}, \frac{m_S^2}{\mu^2}$, $\frac{\mu^2}{m_N^2})$.

\bibliography{darkEDM_bib}

\end{document}